\documentclass[preprint,12pt]{elsarticle}

\usepackage[scale=0.8]{geometry}
\usepackage{amsmath,amssymb,amsfonts}

\usepackage{acro}
\usepackage{xcolor}
\usepackage{multirow}
\usepackage{graphicx}
\usepackage{caption}
\usepackage{booktabs}
\usepackage{url}
\usepackage{lineno}
\usepackage{tcolorbox}
\usepackage{enumitem}

\graphicspath{{Figures/}}

\newcommand{\fw}{0.9\textwidth}

\DeclareAcronym{simo}{short={SIMO}, long={single-input multiple-output}}
\DeclareAcronym{mimo}{short={MIMO}, long={multiple-input multiple-output}}
\DeclareAcronym{shm}{short={SHM}, long={structural health monitoring}}
\DeclareAcronym{pbshm}{short={PBSHM}, long={population-based structural health monitoring}}
\DeclareAcronym{stid}{short={SYSID}, long={system identification}}
\DeclareAcronym{lvv}{short={LVV}, long={Laboratory for Verification and Validation}}
\DeclareAcronym{uos}{short={UoS}, long={University of Sheffield}}
\DeclareAcronym{dstl}{short={DSTL}, long={Defence Science and Technology Laboratory}}
\DeclareAcronym{raf}{short={RAF}, long={Royal Air Force}}
\DeclareAcronym{spo}{short={SPO}, long={sensor placement optimisation}}
\DeclareAcronym{psd}{short={PSD}, long={power spectral density}}
\DeclareAcronym{orpm}{short={ORPM}, long={odd random-phase multisine}}
\DeclareAcronym{rtd}{short={RTD}, long={resistance temperature detector}}
\DeclareAcronym{fbg}{short={FBG}, long={fibre-Bragg grating}}
\DeclareAcronym{sea}{short={SEA}, long={statistical energy analysis}}
\DeclareAcronym{frf}{short={FRF}, long={frequency response function}}

\journal{Structural Health Monitoring}

\begin{document}

\begin{frontmatter}

    \title{Multiple-input, multiple-output modal testing of a Hawk T1A aircraft: A new full-scale dataset for structural health monitoring}

    \author{James Wilson$^{\mathrm{a}}$}
    \author{Max D. Champneys$^{\mathrm{a}}$}
    \author{Matt Tipuric$^{\mathrm{a}}$}
    \author{Robin Mills$^{\mathrm{a}}$}
    \author{David J. Wagg$^{\mathrm{a, b}}$}
    \author{Timothy J. Rogers$^{\mathrm{a}}$}

    \affiliation[1]{
        organization={Dynamics Research Group, University of Sheffield},
        addressline={Western Bank},
        city={Sheffield},
        postcode={S10 2TN},
        country={United Kingdom}}

    \affiliation[2]{
        organization={Alan Turing Institute},
        addressline={96 Euston Road},
        city={London},
        postcode={NW1 2DB},
        country={United Kingdom}}

    \begin{abstract}

        The use of measured vibration data from structures has a long history of enabling the development of methods for inference and monitoring. In particular, applications based on \acl{stid} and \acl{shm} have risen to prominence over recent decades and promise significant benefits when implemented in practice. However, significant challenges remain in the development of these methods. The introduction of realistic, full-scale datasets will be an important contribution to overcoming these challenges. This paper presents a new benchmark dataset capturing the dynamic response of a decommissioned BAE Systems Hawk T1A. The dataset reflects the behaviour of a complex structure with a history of service that can still be tested in controlled laboratory conditions, using a variety of known loading and damage simulation conditions. As such, it provides a key stepping stone between simple laboratory test structures and in-service structures. In this paper, the Hawk structure is described in detail, alongside a comprehensive summary of the experimental work undertaken. Following this, key descriptive highlights of the dataset are presented, before a discussion of the research challenges that the data present. Using the dataset, non-linearity in the structure is demonstrated, as well as the sensitivity of the structure to damage of different types. The dataset is highly applicable to many academic enquiries and additional analysis techniques which will enable further advancement of vibration-based engineering techniques.

    \end{abstract}

    \begin{keyword}
        \acs{mimo}
        \sep Modal testing
        \sep \Acl{shm}
        \sep Hawk T1A
        \sep \Acl{stid}
    \end{keyword}

\end{frontmatter}

\section{Introduction} \label{sec:Introduction}

	The unexpected failure of many types of structures -- including bridges, buildings, aircraft, power infrastructure, and marine vessels -- can be both financially costly and present severe risk to life. The field of \acf{shm} aims to use data collected from the structure to detect the presence of damage, localise its position, categorise its type, assess its extent, and predict the future life of the structure \cite{phd:Ryt:1993, art:Wor:2004, art:Wor:2007, book:Far:2013}.
	
	Vibration-based monitoring of structures is a highly informative process for examining engineering structures which is often used for \ac{shm}. Modal characteristics are particularly useful for inferring structural properties; this is due to their sensitivity to both the global and local physics of the structure, as well as their relatively low dimensionality \cite{book:Far:2013}. In predominantly linear systems, damage (in the form of cracks and material failure) is often detectable as a local loss of stiffness, leading to a shift in the natural frequencies of the system. However, the presence of non-linearity can lead to both false positives and false negatives \cite{art:Hem:2001}. The additional difficulty of discriminating between data from a structure in its normal condition (with potentially variable operating conditions \cite{cross2011cointegration, dervilis2015robust}) and in the presence of structural damageis an ongoing challenge in \ac{shm}. 
	
	The development of robust methods for \ac{shm} requires meaningful, high-quality, openly available benchmark datasets. Furthermore, there is an urgent need to bridge the gap between simple laboratory structures and full-scale structures in service, since many existing datasets are either overly simplistic or highly complex: these are either limited to civil structures (such as the Z24 road bridge \cite{art:Mae:2003} and LUMO lattice mast \cite{art:Wer:2022}), individual components (such as turbine blades \cite{art:Tat:2021, art:Hay:2024}), or bench-top models with comparatively simple geometries \cite{art:Wor:2018, osti_961604}. Bridging the gap between these two levels of complexity this will require datasets which:
	
	\begin{enumerate}
	    \item Are sufficiently realistic to describe in-service engineering structures.
	    \item Are sufficiently complex that system identification is not trivial using current methods.
	    \item Contain data from a number of different conditions (including operating conditions and damage states).
	\end{enumerate}
	
	In this paper the authors present a new benchmark dataset for the purposes of \ac{shm} and \acf{stid}. This dataset consists predominantly of force, acceleration and strain data collected from dynamic testing of a BAE Systems Hawk T1A aircraft. The work significantly extends a previous dataset collected by Haywood-Alexander et al. \cite{art:Hay:2024}, which considered only the starboard wing of the same aircraft. In addition to extending vibration testing of the both the undamaged and pseudo-damaged structure to the entire body of the aircraft, the current dataset also builds on the previous work by including data from \acf{mimo} testing, multi-site pseudo-damage tests, and data from actual damage induced by removing panels from the skin of the structure.
	
	The dataset is freely provided with the intention that it might be used by others to develop new algorithms for structural dynamics. Access is facilitated via a Python interface\footnote{https://github.com/MDCHAMP/hawk-data} developed by the authors for both this dataset and the previous dataset by Haywood-Alexander et al. \cite{art:Hay:2024}.
	
	The remainder of this paper is structured as follows: firstly, the Hawk structure is described in detail. Secondly, the experimental work is described, including discussions of the testing regime, setup and hardware, and acquisition and control. A further section describes the dataset itself. Finally, the challenges that this new dataset presents in a \ac{shm} context are explored, including \ac{stid} and \ac{shm}. The principal contributions of this work are as follows:
	
	\begin{itemize}
	    \item A full-scale \ac{mimo} vibration test of the BAE Systems Hawk T1A aircraft is presented as a new benchmark dataset.
	    \item The dataset includes broadband excitation data from five locations on the Hawk, which was instrumented with over 140 sensors.
	    \item The dataset comprises over 200 test conditions including different excitation signals and amplitudes, as well as comprehensive damage simulations.
	    \item The entire dataset is made freely available alongside with a convenient Python interface that enables users to work with a subset of the data as required.
	\end{itemize}

\section{Structure} \label{}

	\begin{figure}
	    \centering
	    \includegraphics[width=\fw]{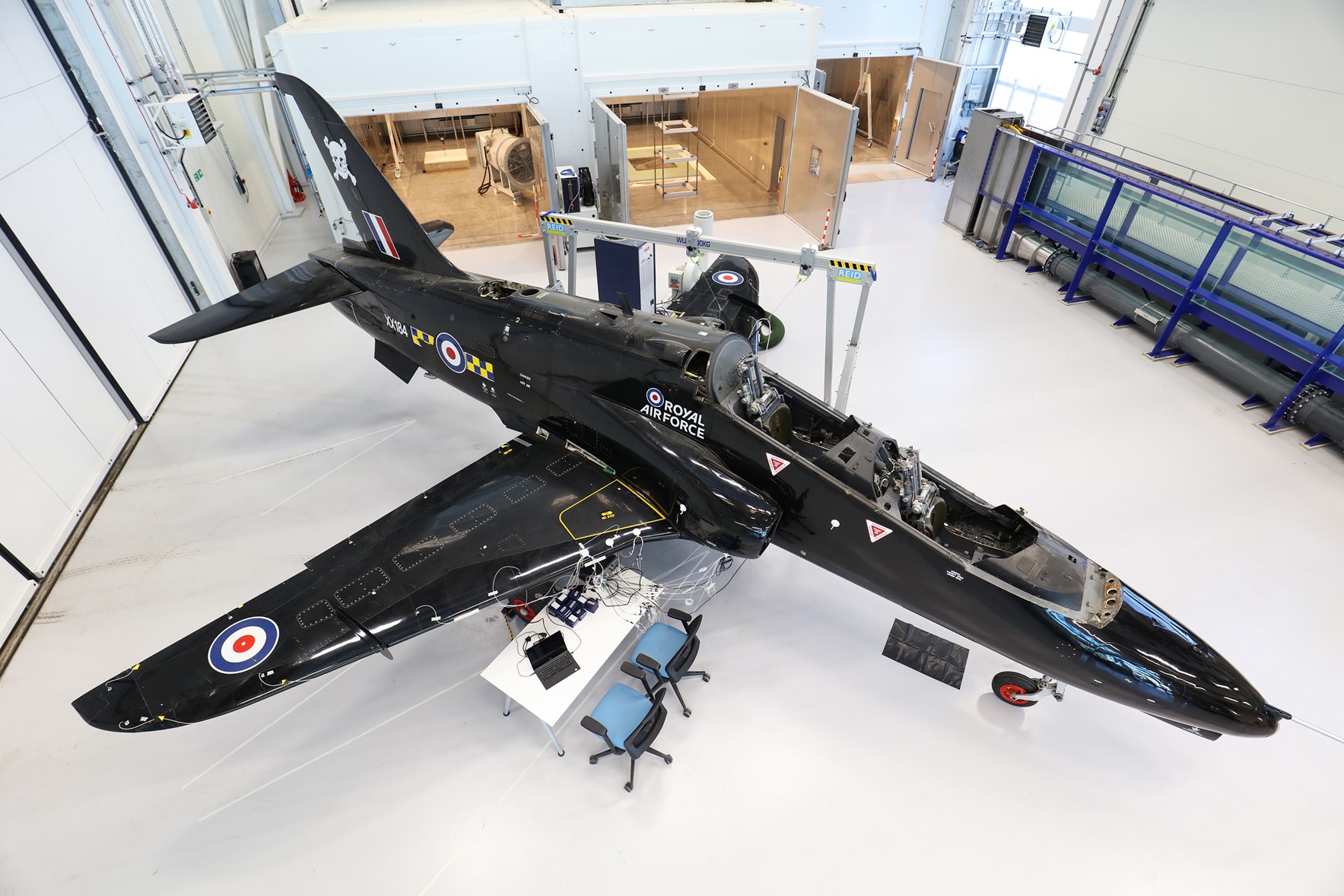}
	    \caption{The Hawk at the \acl{lvv}.}
	    \label{fig:Hawk_clean}
	\end{figure}
	
	The structure presented in this paper is a BAE Systems Hawk T1A (`the Hawk') that was donated to the University of Sheffield by the Defence Science and Technology Laboratory. The Hawk is housed at the \acf{lvv} within the University; it is a decommissioned aircraft previously used for advanced training by the British Royal Air Force. Structurally, the airframe of the aircraft is largely complete. However, the engine, certain wing flaps, the cockpit canopy and various other components have been removed. The aircraft is positioned resting on its wheels on the ground. A full set of technical drawings is not available.
	
\section{Experimental work} \label{}

\begin{figure}
		    \centering
		    \includegraphics[width=0.8\textwidth]{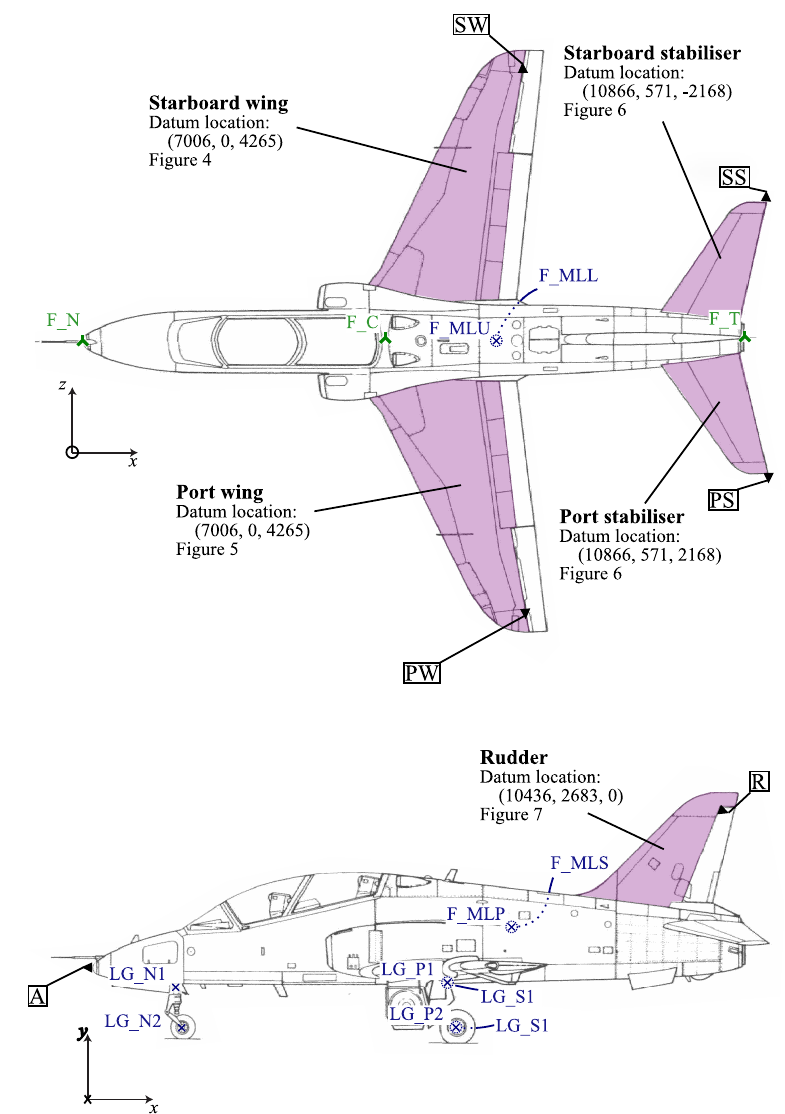}
		    \caption{Summary sensor placement sketch for the Hawk, showing the locations of the nearside (\includegraphics{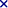}), farside (\includegraphics{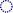}) and triaxial (\includegraphics{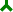}) accelerometers  on the main chassis. Also shown are the locations of the datums (SW, SS, PS, and PW) relative to the global datum, A. The substructures highlighted in magenta are detailed in Figures \ref{fig:schematic_swing}-\ref{fig:schematic_rudder}. The diagram is not shown to scale -- for precise locations, refer to the data set \cite{data:Wil:2024}.}
		    \label{fig:schematic_hawk_summary}
		\end{figure}

\begin{figure}
		    \centering
		    \includegraphics[width=\fw]{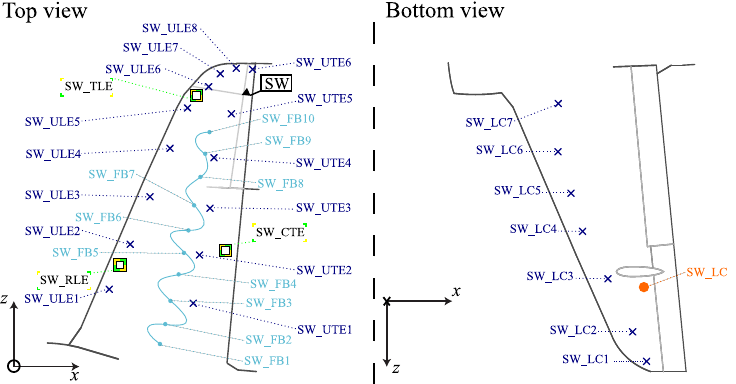}
		    \caption{Schematic diagram for the starboard wing, showing the location of the accelerometers (\includegraphics{icon_accelerometer}), \ac{fbg} (\includegraphics{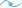}), shakers (\includegraphics{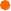}), and pseudo-damage (\includegraphics{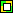}). Not to scale.}
		    \label{fig:schematic_swing}
		\end{figure}

\begin{figure}
		    \centering
		    \includegraphics[width=\fw]{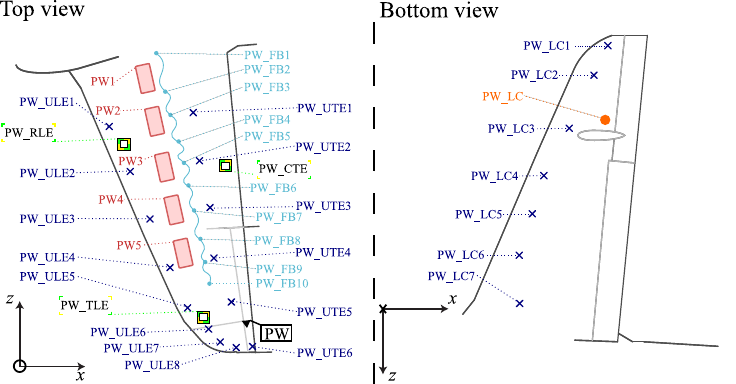}
		    \caption{Schematic diagram for the port wing, showing the location of the accelerometers (\includegraphics{icon_accelerometer}), \ac{fbg} (\includegraphics{icon_FBG}), shakers (\includegraphics{icon_shaker}), pseudo-damage (\includegraphics{icon_damage}), and removed panels (\includegraphics{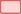}). Not to scale.}
		    \label{fig:schematic_pwing}
		\end{figure}

\begin{figure}
		    \centering
		    \includegraphics[width=\fw]{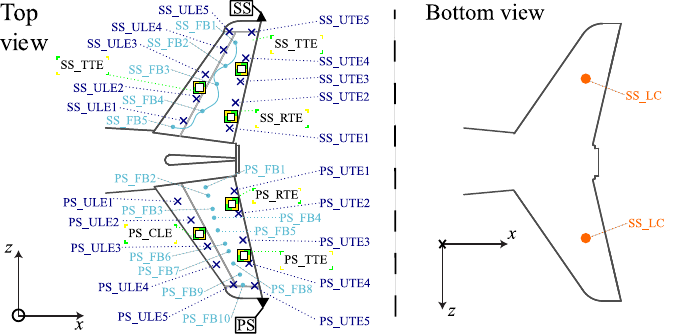}
		    \caption{Schematic diagram for the stabilisers, showing the location of the accelerometers (\includegraphics{icon_accelerometer}), \ac{fbg} (\includegraphics{icon_FBG}), shakers (\includegraphics{icon_shaker}), and pseudo-damage (\includegraphics{icon_damage}). Not to scale.}
		    \label{fig:schematic_stabilisers}
		\end{figure}

	Details of the experimental work are presented in this section. Firstly, the experimental setup and hardware are described, followed by details of the testing regime. Finally, the acquisition and control of the input force are discussed. A full set of schematic diagrams summarising the placement of accelerometers, shakers, \ac{fbg} strain gauges and damage locations are given in Figures \ref{fig:schematic_hawk_summary}-\ref{fig:schematic_rudder} to complement this section.

	\subsection{Setup and hardware}
	
\begin{figure}
		    \centering
		    \includegraphics[width=\fw]{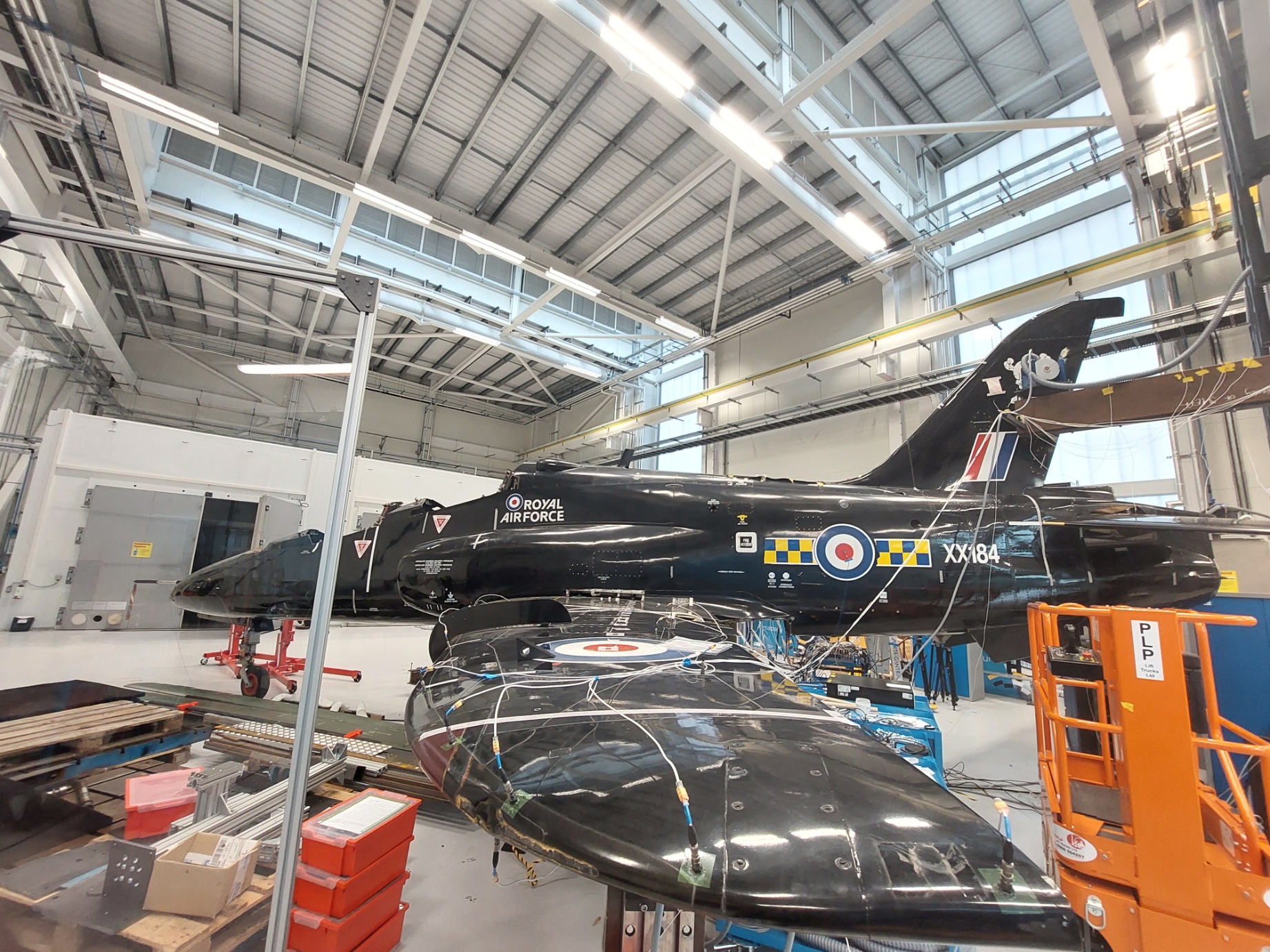}
		    \caption{The port wing of the Hawk during instrumentation.}
		    \label{fig:Hawk_wing}
		\end{figure}

		Five types of sensor were used in these tests: accelerometers, \ac{fbg} strain gauges, force transducers, a \ac{rtd} and a microphone. The force transducers measured the excitation forces from the shakers, the accelerometers recorded the acceleration response and the strain gauges recorded the strain. The ambient temperatures were recorded using the \ac{rtd} and a triaxial accelerometer recorded ground vibrations. All of these measurements were recorded in the time domain. Ambient noise was also recorded using a microphone; for privacy reasons this was not recorded in the time domain -- a Fourier transform of these data was recorded instead. An image of the Hawk during instrumentation is shown in Figure \ref{fig:Hawk_wing}. 
		
		The electrodynamic shakers were placed to excite each wing, each stabiliser and the rudder. This was to ensure that a strong dynamic response could be recorded across the whole structure. Their precise location was guided by limitations of the structure (such as the need to avoid exciting near the landing gear). For excitation, Tira TV 51140-MOSP modal shakers were used in conjunction with BAA 1000 amplifiers. The force transducers were PCB Piezotronics 208C02s, with a nominal sensitivity of 11241mV/kN.
		

		To measure the structural response, PCB Piezotronics accelerometers were used with a nominal sensitivity of 10mV/g. When converting voltage readings to acceleration values, the sensor-specific sensitivities (contained within the metadata \cite{data:Wil:2024}) were used. A set of triaxial accelerometers were placed at the nose, cockpit and tail of the Hawk fuselage; PCB Piezotronics accelerometers were used with a nominal sensitivity of 100mV/g. A set of four uniaxial PCB Piezotronics accelerometers, also with a nominal sensitivity of 100mV/g, were placed on the fuselage midpoint on the top, bottom, port and starboard sides. Pairs of these sensors were also placed on each landing gear; one above the hydraulic suspension and one on the axle of the wheel. A total of 85 accelerometers were placed on the Hawk.
		
		\ac{fbg} strain gauges were mounted along the length of each wing, as well as along each stabiliser and up the rudder. Ten measurement locations were placed on each wing, and on the port stabiliser. A single \ac{fbg} strain gauge was shared between the rudder and the starboard stabiliser, meaning that five measurement locations were available on each of these substructures. The precision of the strain gauges was 1 microstrain and their measurement length was 10mm; on a large surface such as the Hawk skin these can effectively be considered as point measurements.
		
		One PCB Piezotronics triaxial accelerometer, with a nominal sensitivity of 100mV/g, was placed on the ground. A microphone was placed under the fuselage of the Hawk to record any significant ambient noise. The microphone was a GRAS 46AE with a precision of 0.08dB. The \ac{rtd} used was an RS PRO PT100 sensor, and was accurate to $0.1^{\circ}$C. Exact sensitivities of all sensors are available as part of the dataset record at \cite{data:Wil:2024}.

\begin{figure}
		    \centering
		    \includegraphics[width=\fw]{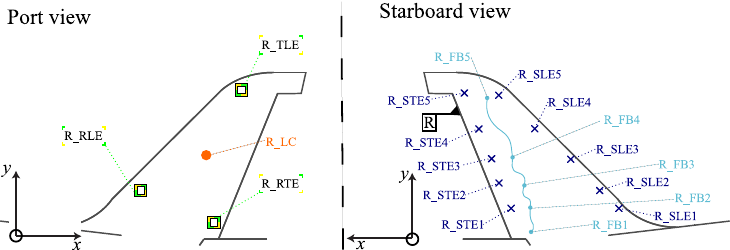}
		    \caption{Schematic diagram for the rudder, showing the location of the accelerometers (\includegraphics{icon_accelerometer}), \ac{fbg} (\includegraphics{icon_FBG}), shaker (\includegraphics{icon_shaker}), and pseudo-damage (\includegraphics{icon_damage}). Not to scale.}
		    \label{fig:schematic_rudder}
		\end{figure}

		The coordinate locations for the sensors were measured on the structure itself where possible; these were measured from local reference points on each surface. Despite some difficulties in measuring distances across curved surfaces, the location of these points are sufficiently accurate for \ac{stid}. The global datum was taken from the tip of the nose (labelled A in Figure \ref{fig:schematic_rudder}) of the Hawk. The global measurements, including the coordinates of the various reference points, were measured from drawings of the Hawk, which were scaled to a particular reference length on the real structure. The coordinates shared as part of the dataset are intended for use in visualisation of the structure only; more advanced measurement techniques would be required to derive a set of fully accurate coordinates. Most sensors on the structure were aligned along a particular coordinate axis. Wherever there was a significant deviation in this direction, this is noted in the data record. All sensor coordinate data is available in \cite{data:Wil:2024}.
		
		The landing gear tyres were inflated to a pressure of 9 bar prior to testing. Following a period of downtime in the \ac{lvv} from 10$^{\text{th}}$--25$^{\text{th}}$ August 2023, the tyres were repressurised to this value on 29$^{\text{th}}$ August 2023.
		
	\subsection{Testing regime}
	
		The experimental work was carried in five phases. All test phases were carried out in \ac{mimo} configuration, with the structure excited by electrodynamic shakers attached to each wing and stabiliser and at the rudder.
		
		The first phase concerned the acquisition of data to fully describe the normal, undamaged condition of the Hawk. The aircraft was excited by signals with a range of amplitudes, enabling the identification of any amplitude-dependent non-linearities. Input excitation signals were designed as white noise and pink noise signals. For all tests, the excitation bandwidth was 5--256Hz and the sampling rate was 2048Hz.
		
		Following the undamaged tests, the Hawk was tested in a range of single-site damage scenarios. Pseudo-damage was introduced to the structure by the addition of weights of varying mass at a range of locations \cite{art:Pap:2010}. The masses were attached to the surface of the Hawk by suction cup. Details of the masses used are given in Table \ref{tab:masses}. Damage locations are presented in the schematic Figures \ref{fig:schematic_swing}-\ref{fig:schematic_rudder}. White noise excitation was used for these tests, where the excitation bandwidth was 5--256Hz and the sampling rate was 2048Hz. A range of excitation amplitudes were used, corresponding to those used in the first testing phase.
		
		\begin{table}
		    \centering
		    \begin{tabular}{c|c}
		        Name & Mass (g) \\\hline
		        M1   & 254.3    \\
		        M2   & 616.8    \\
		        M3   & 916.8    \\
		        M6   & 350.3    \\
		        M7   & 64.4     \\
		    \end{tabular}
		    \caption{Details of the masses used for approximating damage in the Hawk}
		    \label{tab:masses}
		\end{table}
		
		The third phase of testing entailed excitation of the structure in its undamaged condition using \ac{orpm} excitation. This is a popular input design method to enable non-linear \acf{stid} in the frequency domain. The excitation profile contains a random selection of odd-frequency sine waves; leakage into the response of other frequency content can then be treated as an indicator of non-linearity in the structure \cite{art:Csu:2022}. The \ac{orpm} tests were carried out with the Hawk in its undamaged condition, and also with some single-site mass additions. Initial \ac{orpm} tests were conducted in a \ac{simo} configuration considering each input location separately, with following tests carried out as previously in a \ac{mimo} configuration.
		
		The fourth phase of testing investigated the response of the Hawk to multisite damage. The structure was excited using white noise signals in the range 5--256Hz and the sampling rate was again 2048Hz. The same masses and locations as were used in the single-site damage tests were employed here. Up to three masses were positioned on the structure at any one time.
		
		The final phase of testing involved a series of actual damage-state tests where surface panels were removed from the port wing of the Hawk. The panels were arranged in a line along the length of the wing; five panels were identified for the tests and were labelled PW1--PW5. The panel removal tests were carried out at the end of the testing to avoid the impact that the removal and reattachment of any panels may have on the response of the structure in other tests. Only one panel was removed at any time, and for each scenario the structure was excited with white noise signals at three different amplitude levels.
		
		The full experimental campaign comprised 216 individual tests. For the sake of brevity, the full list is not tabulated here. The interested reader is directed to the dataset at \cite{data:Wil:2024} for the full downloadable tabular record of test conditions.
	
	\subsection{Acquisition and control}
	
		The majority of testing on the Hawk was carried out using random excitation, with white noise and pink noise signals used. The spectral densities of the input signals were pre-determined using `breakpoint' tables. These tables define a rectangular or trapezoidal spectral density profile (for white and pink noise respectively) in the frequency domain for each shaker. The highest input amplitude was limited by the maximum safe level of excitation that could be applied to the Hawk via stinger, and the testing amplitudes were reduced linearly from these upper bounds. In total, five excitation levels are defined for both white and pink noise excitations.
		
		Sample breakpoint tables for white noise and pink noise excitation are given in Table \ref{tab:white_noise_1_body} and Table \ref{tab:pink_noise_1_body}. The breakpoint tables were resampled 8192 times to generate an amplitude `mask' across the full bandwidth (up to 2048Hz). For a full description of all breakpoint tables and input signal metadata, the reader is directed to the dataset record \cite{data:Wil:2024}.

		\begin{table}
		    \centering
		    \begin{tabular}{c|c}
		        Frequency (Hz) & \Acl{psd} (N$^2$/Hz) \\ \hline
		        0              & 0             \\
		        4              & 0             \\
		        5              & 0.02          \\
		        256            & 0.02          \\
		    \end{tabular}
		    \caption{The breakpoint table for the excitation signal `white\_noise\_1'}
		    \label{tab:white_noise_1_body}
		\end{table}
		
		\begin{table}
		    \centering
		    \begin{tabular}{c|c}
		        Frequency (Hz) & \Acl{psd} (N$^2$/Hz) \\\hline
		        0              & 0             \\
		        4              & 0             \\
		        5              & 0.12          \\
		        256            & 0.02          \\
		    \end{tabular}
		    \caption{The breakpoint table for the excitation signal `pink\_noise\_1'}
		    \label{tab:pink_noise_1_body}
		\end{table}
		
		It should be noted here that the signals used are generally lower in amplitude than those used for the starboard-wing tests by Haywood-Alexander et al. \cite{art:Hay:2024}. This is due to the \ac{mimo} nature of this testing campaign. Overall, a greater total excitation energy is supplied compared to a single-shaker setup. In this test campaign, excitation was additionally applied to substructures of the Hawk tailplane (stabilisers, rudder) that are likely to have a dramatically lower stiffness than the wings, requiring less input energy for the same signal-to-noise ratio.
		
		The time-domain drive signals for each input shaker were generated by an iterative procedure. For each of the five shakers, a unique drive signal was created by generating a random summation of sine waves, each of which was assigned a random phase. The initial amplitude mask (in the frequency domain) could then be generated by multiplying the breakpoint table by an initial value of 0.01, which provided a conservative point at which to begin iterating the drive signals. Each shaker was driven with its corresponding drive signal and the output measured at the load cell was compared to the target \ac{psd} defined by the breakpoint table. The quotient of the two was then used as a multiplier to update the drive signal for the next iteration. This process was iterated until the error between the measured and target \ac{psd}s met a defined tolerance -- set to $4\%$ for these tests -- for all shakers, at which point the spectral gain mask was stored for each shaker.
		
		The drive signals were then generated for each shaker according to the spectral gain masks. The spectral gain masks were resampled in the frequency domain to differing lengths by linear interpolation to allow for variable test periods to be specified. This would introduce some error on the desired excitation, which could be captured by interrogating the measured excitation signals for each test. New spectral drive masks were generated if the structure was modified between tests (such as by adding a pseudo-damage mass) or at the beginning of each test session to account for the effects of any variations on the structure. 
		
		For the \ac{orpm} tests, a mask of ones and zeros was generated to accept the odd spectral content and reject the even content. This was used as an additional multiplier when generating the drive files. 50\% of the odd frequency content was retained for each iteration of the \ac{orpm} tests.

\section{Dataset} \label{}

	Overall, 216 tests were conducted during the campaign, each comprised of 139 individual sensor channels with an average of 10 repeats per test. The raw size of the collected dataset exceeds 500GB, far in excess of most users' capacity to load it into memory.
	
	The Hawk full-structure dataset is packaged in the widely used \emph{hierarchical data format} (.hdf5). The .hdf5 format allows data and metadata to be stored concurrently and facilitates lossless data compression. Many software packages are available for opening and exploring .hdf5 files and programmatic access is implemented in most popular programming languages.
	
	In order to further facilitate simple and practical access to the data, a simple interface has been created in the Python language. The Python interface\footnote{Available at github.com/MDCHAMP/hawk-data} automatically handles the downloading and storage of the Hawk data from the centralised repository \cite{data:Wil:2024} ensuring that the data is up-to-date and avoiding generation loss by repeated sharing. Data is accessed programmatically meaning that only data that is required by the user is downloaded. The API first checks for local versions of the files and, if they do not exist, accesses the data from the central repository, comparing the MD5 hash to ensure the data is up-to-date. Note that use of the Python interface is not required in order to obtain the data, it is only provided to simplify the process for users who require such functionality.
	
	In order to serve only the data that is required to the user, the whole dataset is divided into discrete units. This is achieved by leveraging the hierarchical nature of the .hdf5 file format and using external links to connect a number of separate .hdf5 files to a single .hdf5 header file. Data files are atomised on a per-test basis, including all repeats and metadata. Dividing the dataset in this way allows it to be distributed in single-test files, each around 1GB in size (rather than a single compressed record of around 250GB). Packaging the data in this way drastically lowers the required disk footprint for users who only require a subset of the test conditions for benchmarking and analysis.
	
	Additionally (as in \cite{art:Hay:2024}), the dataset has been compiled in a \emph{self-describing} format. Self-describing datasets package pertinent metadata alongside raw sensor output data in order to create a data record that does not require the user to have detailed knowledge of the test campaigns in order to use the data. In the context of the Hawk full structure test metadata (i.e. signal name, sensor, sensor type, sensitivity, sensor location, units and test conditions) are stored alongside each channel in a hierarchical format. For a more complete description of which metadata are available the interested reader is directed to the usage examples on the data repository \cite{data:Wil:2024}.
	
	

\section{Dataset challenges} \label{sec:dataset_challenges}

	In Section \ref{sec:Introduction}, the authors argued that bridging the gap between \ac{shm} of laboratory and full-scale structures requires the use of realistic, complex datasets with data from  a number of different conditions. It is hoped that the dataset presented in this paper will be of interest to a wide range of researchers with a shared interest in understanding full-scale dynamic systems. In this section, some open research challenges that the dataset highlights are discussed. These challenges pertain to the practicalities of working with large datasets, \ac{stid} of a complex structure, and \ac{shm}. Following each of these subsections, the key challenges are enumerated for ease of reference. 
	
	In order to visualise some salient aspects of the Hawk dataset, several \ac{frf}s are computed here in Figures \ref{fig:all_accs}-\ref{fig:panel_PW}. However even these simple analyses represent a non-trivial challenge for the Hawk dataset. Well-established methodologies can present challenges on large-scale datasets. In the context of the Hawk aircraft, a natural way to characterise the (predominantly) linear dynamics is through the lens of modal analysis \cite{Ewins2009}. In the \ac{mimo} setting, the \ac{frf} matrix can be estimated from the measured input-output data. For a linear system in the frequency domain, one has
	
	\begin{equation}
	    Y(\omega) = H(\omega) X(\omega) \label{eq:Y}
	\end{equation}
	
	\noindent where $H$ is the \ac{mimo} \ac{frf}, and  $X$ and $Y$ are the Fourier transforms of the input and output data respectively. Post-multiplying the above by the conjugate transpose of $X$ and taking expectations gives
	
	\begin{equation}
	    S_{xy}(\omega) = H(\omega) S_{xx}(\omega) \label{eq:Sxy}
	\end{equation}
	
	\noindent where $S_{xy}$ is the cross-spectral density of the input with the output and $S_{xx}$ is the auto-spectral density of the output. The linear Equation \ref{eq:Sxy} can thus be solved for $H$ at each frequency line. However, a naïve solution to the above for many input and output channels is likely to be poorly conditioned numerically and lead to corruption and artefacts in the in identified \ac{frf}. In the present work, \ac{frf}s for visualisation are computed via a short-time Fourier transform (STFT) with a Hann window, a segment length of 16384 and an overlap of 8192. The resulting linear system is solved in the least-squares sense by taking the inner product over the number of segments in the STFT. 
	
	\subsection{Large datasets}
	
	This dataset represents a comprehensive set of testing which may be carried out on a structural dynamic system. Inevitably, the testing of full-scale structures across a range of conditions gives rise to relatively large datasets, which may not be easily analysed on a local computer due to limitations in memory. As data-driven approaches to \ac{shm} become adopted more widely in industrial settings, the necessity of handling datasets of this scale will grow. In the context of an academic research environment it is worth highlighting that as methodologies are developed, this context should be taken into account. The practicality of using already-established algorithms in the context of large datasets also needs to be evaluated.
		
	Additionally, as the size of these datasets increases, reliance on \emph{ad hoc} human interpretation of the data becomes harder and the need for further levels of automation in analysis will increase. For example, inspecting Figure \ref{fig:all_accs}, it is clear that the complexity of a structure such as the Hawk aircraft is very high and direct observations (even from a skilled operator) may be limited. In order for these techniques to be practical in industrial settings, operators need to be able to check, in a realistic timescale, that the data is being collected as expected. The development of practical workflows and the implementation of automated checks present an open challenge. The key challenges can thus be summarised:
	
	\begin{tcolorbox}
	    \begin{enumerate}[label={Challenge \arabic*.}, leftmargin=2.5cm, listparindent=-\leftmargin]
	        \item{Using existing and new \ac{stid} algorithms on large datasets in realistic time frames.}
	        \item{Aiding (or reducing the reliance on) operator interpretation when collecting large datasets.}
	    \end{enumerate}
	\end{tcolorbox}
	
	\subsection{\ac{stid}}
	
	It is envisioned that the Hawk dataset will provide a meaningful testbed for identification algorithms in the context of structural dynamics. For such dynamics analyses, it is often necessary to determine the underlying properties of the dynamic system (for example, a representative set of governing differential equations) from the available data. This dataset presents a number of challenges to researchers in this area.
	
	It is difficult to develop accurate models of large, built-up structures (such as this aircraft) owing to the richness of the dynamics which is captured. The dataset itself is generated using \ac{mimo} tests, which only increase the quantity of data in comparison to \ac{simo},  and so can require more care in the signal processing.  
	
	In dynamical systems theory it is generally assumed that a system is governed by a set of unknown partial differential equations which cannot be directly determined, and that have complex and unknown boundary conditions. The structure comprises components and subsystems, each of which have varying dynamic properties. These parts are assembled into the full structure in a manner which requires modelling of jointed interactions --- an open area of fundamental research \cite{gaul1997nonlinear, oldfield2005simplified, balaji2020traction}.

\begin{figure}
	    \centering
	    \includegraphics[width=\fw]{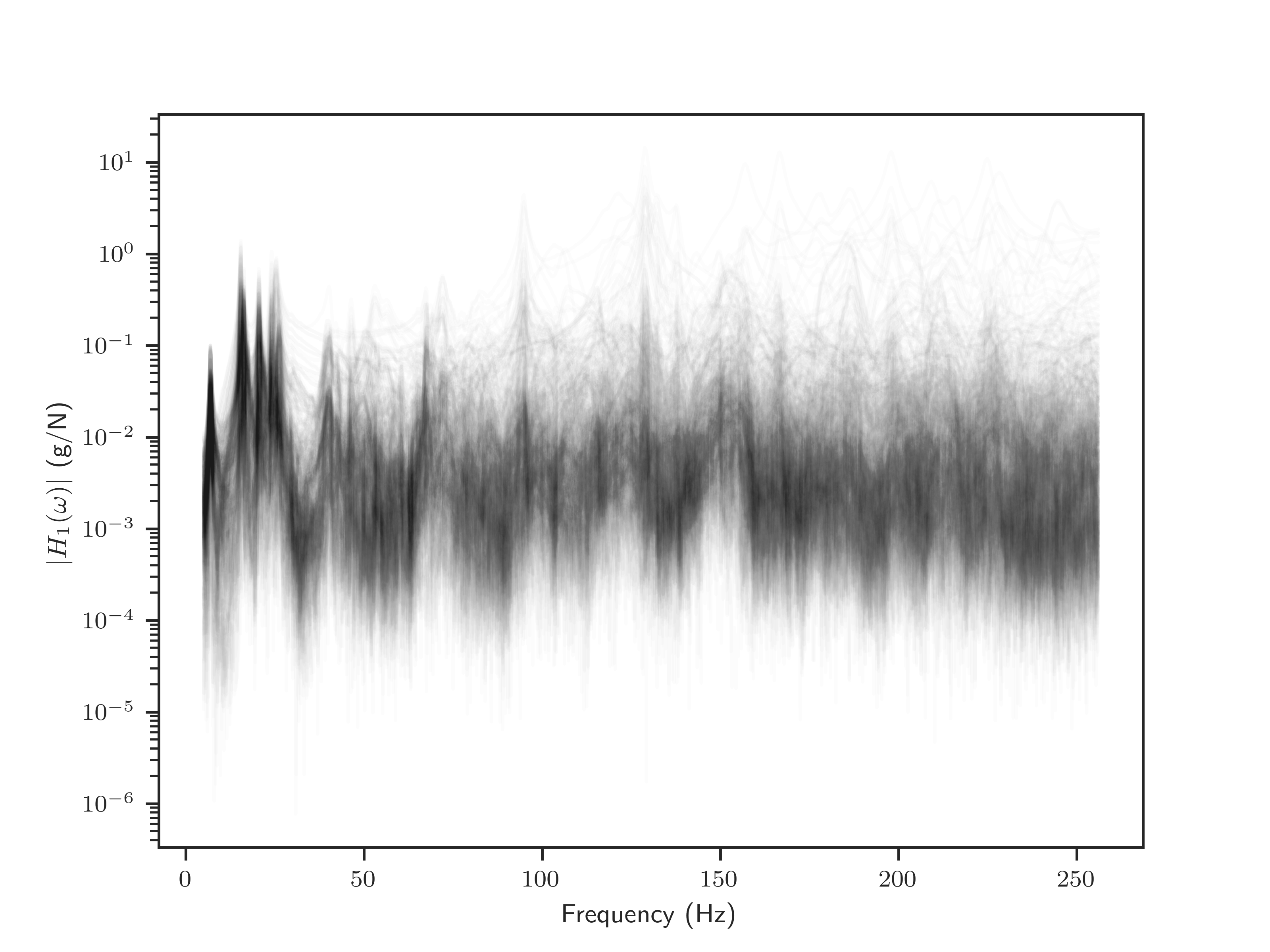}
	    \caption{All accelerometer FRF magnitudes superimposed for the white noise healthy state test at amplitude level 5.}
	    \label{fig:all_accs}
	\end{figure}

\begin{figure}
	    \centering
	    \includegraphics[width=\fw]{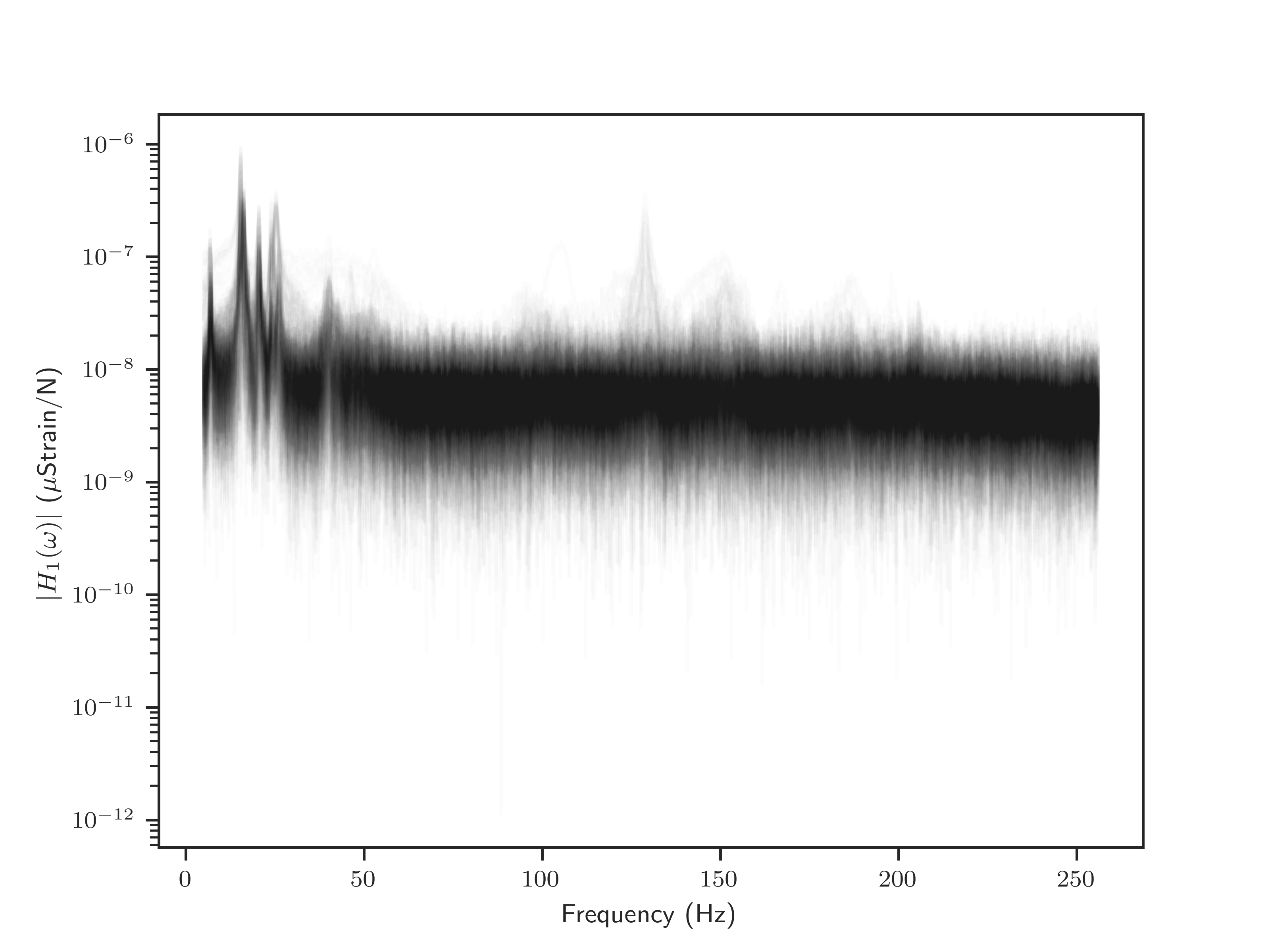}
	    \caption{All \ac{fbg} strain FRF magnitudes superimposed for the white noise healthy state test at amplitude level 5.}
	    \label{fig:all_strain}
	\end{figure}

	Figures \ref{fig:all_accs} and \ref{fig:all_strain} show the combined data recorded from the structure to interrogate the from of the \ac{frf}s for the acceleration and strain sensors respectively. While behaviour can be distinguished in the \ac{frf}s which may be attributed to global modal behaviour in the low frequency region, the full picture is far more complicated. Since the structure contains many thin panels, a global modal behaviour is difficult to resolve. It can therefore be expected that alternative methods that combine deterministic and statistical analyses such as hybrid \ac{sea} \cite{shorter2005vibro}, will be required.
	
\begin{figure}
	    \centering
	    \includegraphics[width=\fw]{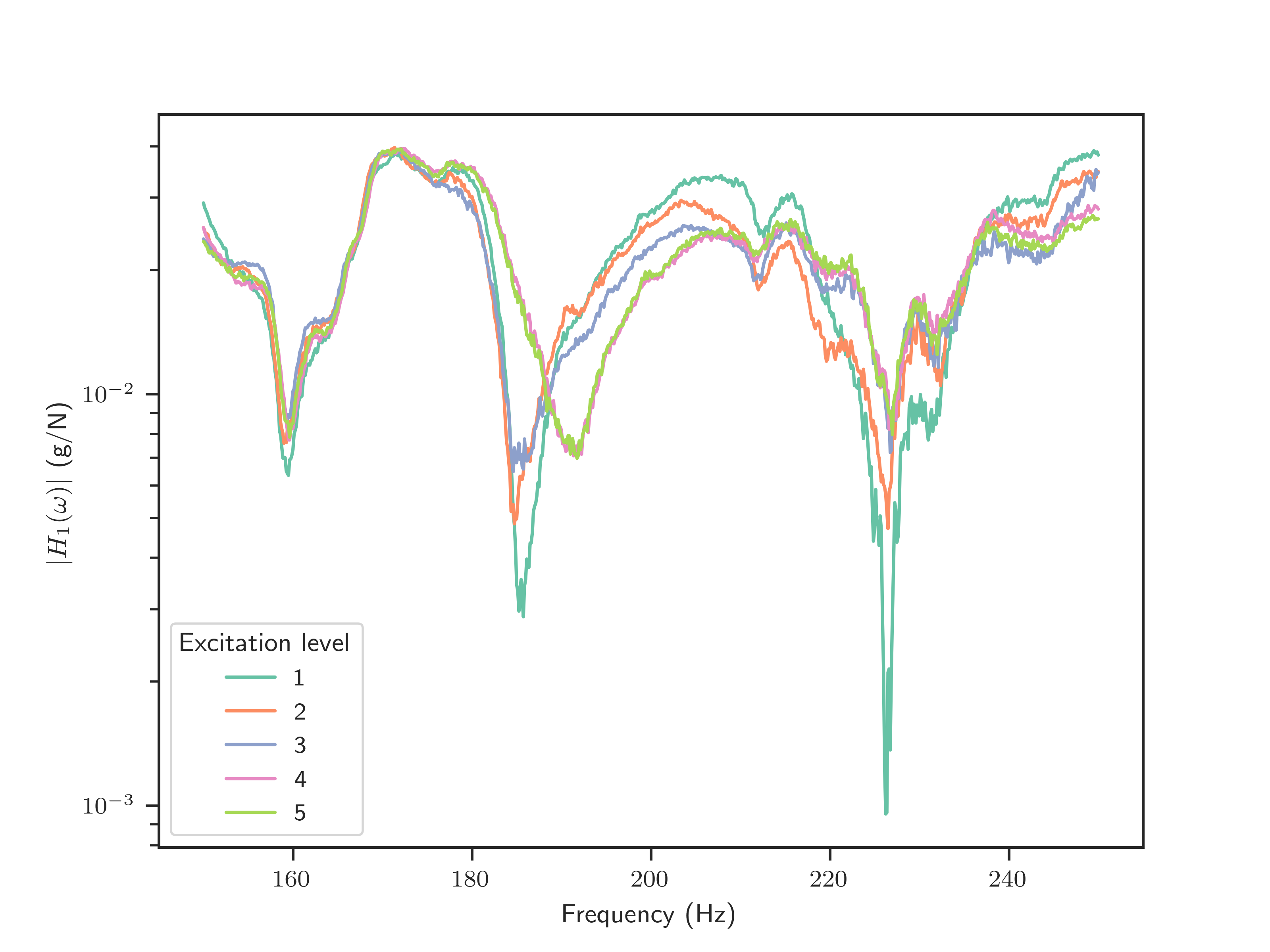}
	    \caption{Cropped FRF magnitude from starboard wing shaker (SW\_LC) to lower starboard landing gear (LG\_S2) depicting non-linear hardening.}
	    \label{fig:nonlin_SLG}
	\end{figure}

	Figure \ref{fig:nonlin_SLG} depicts a subset of the frequency lines from the \ac{frf} of the starboard wing shaker to the lower section of the landing gear on the starboard wing. Here, a further challenge becomes apparent. The \ac{frf} in certain regions of the frequency range can be seen to exhibit strong dependence on the excitation amplitude. This dependence is one indicator that the dynamics of the structure are not well approximated by linear theory. The behaviour observed in \ref{fig:nonlin_SLG} shows some drift resonant peaks with increases in amplitude which may be associated with non-linear hardening behaviour in those resonances (and anti-resonances). The presence of non-linear dynamics in aerospace structures is well known in the literature, for example see a previously published dataset \cite{noel2017f} where testing is carried out on an F-16 aircraft. The identification of non-linear systems with many degrees of freedom remains a challenge and this dataset adds to the utility of \cite{noel2017f} by providing a \ac{mimo} test case and an extended sensor network covering the full structure. It should be noted that the complexities previously discussed regarding the mid-frequency region continue and may require the utilisation of non-linear forms of \ac{sea}, e.g. \cite{spelman2015statistical}.
	
	One common purpose of system identification is to update or validate existing physics-based models of the structure of interest; model calibration and validation are critical to ensuring the accuracy and reliability of a predictive model and model updating is applicable as a damage inference problem. Uncertainty quantification for physics-based models, particularly in the face of epistemic uncertainty as in this dataset \cite{book:Far:2013, phd:Bar:2010}, is a major challenge. The lack of a high-quality model numerical model, as is common with legacy infrastructure, increases the applicability of this challenge.
	
	In the context of a built-up, multi-component structure such as the Hawk, the idea of hierarchical validation offers potential solutions to a range of issues concerned with model validation \cite{phd:Gar:2018, art:Wil:2023}, where relevant challenges would include how to combine datasets collected on parts and subsystems of a more complex overall object. There is an opportunity to investigate model updating in this hierarchical context by using this dataset in combination with the previously presented starboard wing data \cite{art:Hay:2024}.

	The challenges relating to the Hawk dataset in \ac{stid} are summarised below:
	
	
	
	\begin{tcolorbox}
	    \begin{enumerate}[label={Challenge \arabic*.}, leftmargin=3cm, listparindent=-\leftmargin, resume*]
	        \item{Developing accurate models of large, built up structures using MIMO data.}
	        \item{Modelling the dynamic properties of structures with jointed subsystems.}
	        \item{The identification of global modal behaviour.}
	        \item{The identification of non-linearities within the system.}
	        \item{Uncertainty quantification and physics-based model updating in the face of epistemic uncertainty.}
	        \item{Hierarchical validation, in conjunction with the previous dataset.}
	    \end{enumerate}
	\end{tcolorbox}
	
	\subsection{\ac{shm}}
	
	\begin{figure}
	    \centering
	    \includegraphics[width=\fw]{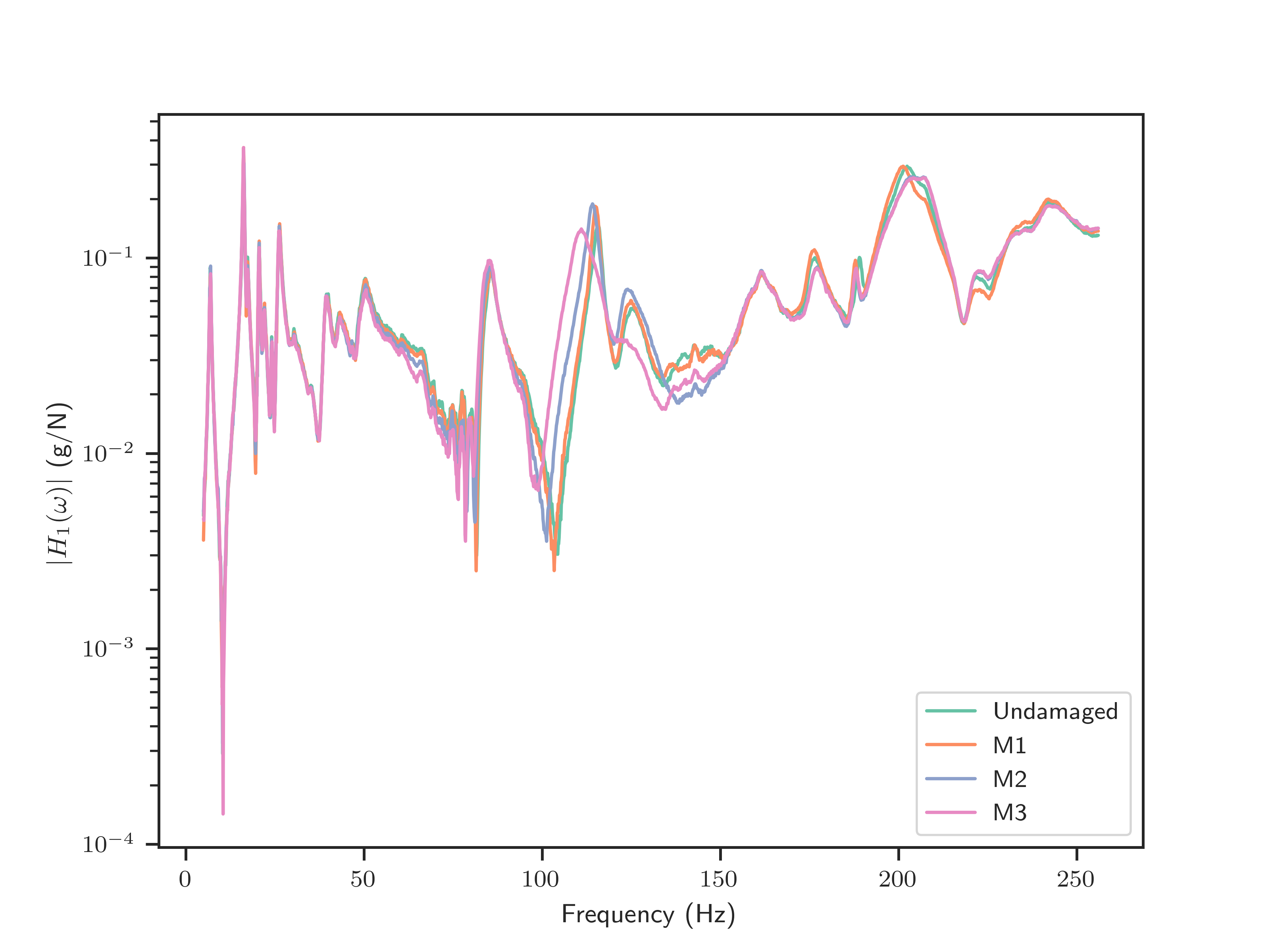}
	    \caption{FRF magnitude from port wing shaker to PS\_ULE5 in the presence of damage at PW\_TLE (amplitude level 5).}
	    \label{fig:dmg_PW}
	\end{figure}
	
	\begin{figure}
	    \centering
	    \includegraphics[width=\fw]{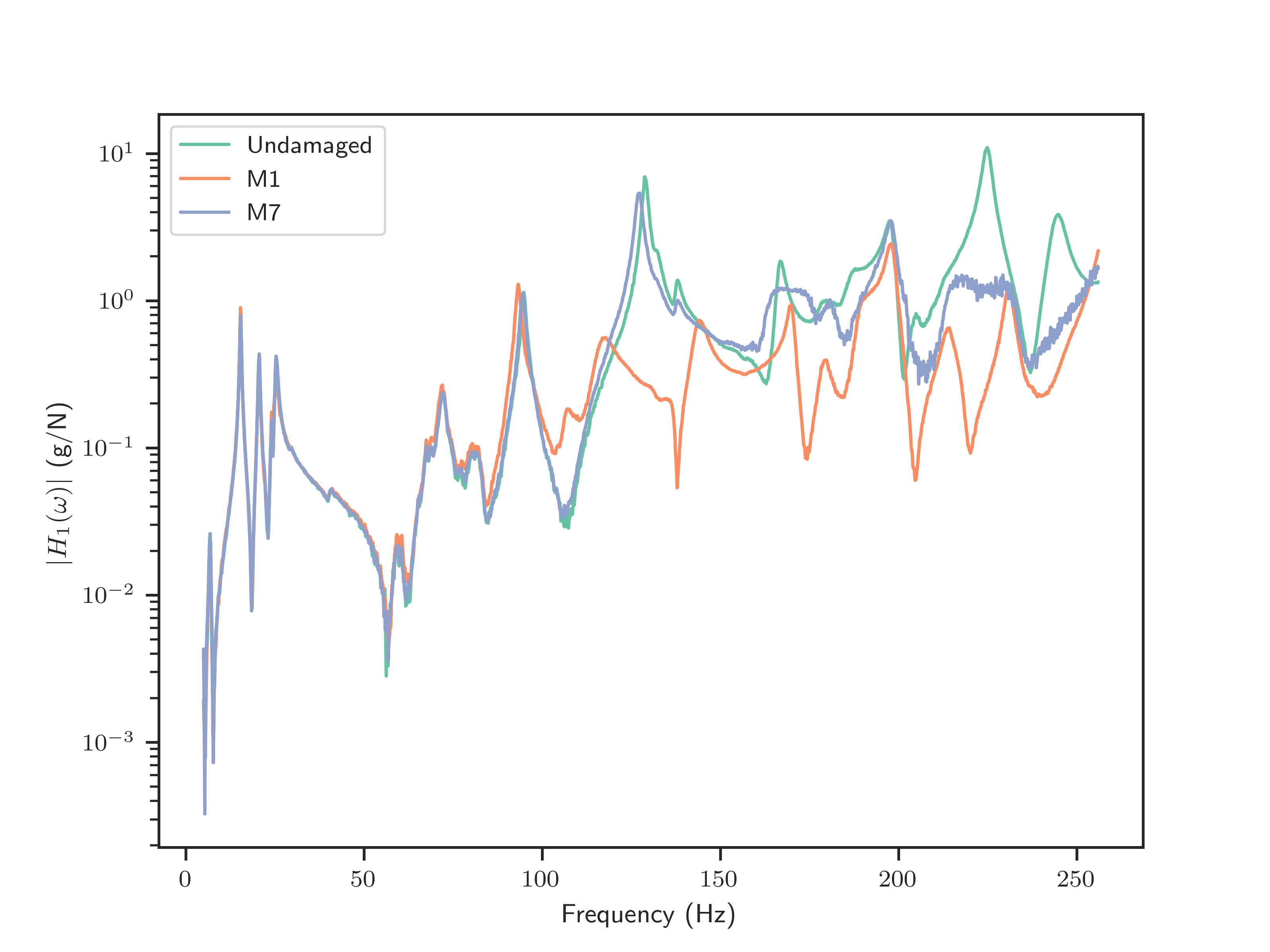}
	    \caption{FRF magnitude from starboard stabiliser shaker to SS\_UTE3 in the presence of damage at SS\_TTE (amplitude level 5).}
	    \label{fig:dmg_SS}
	\end{figure}
	
	\begin{figure}
	    \centering
	    \includegraphics[width=\fw]{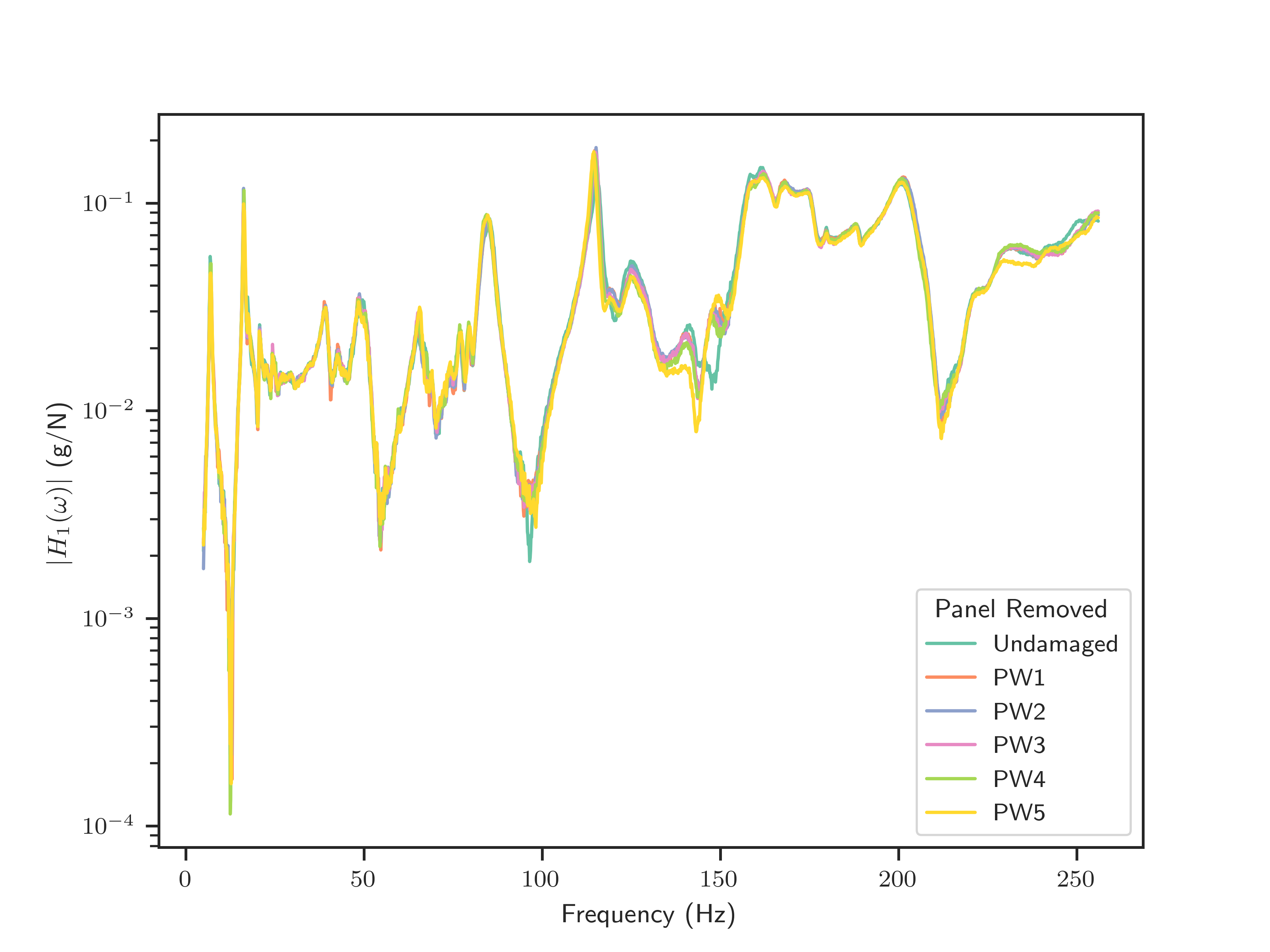}
	    \caption{FRF magnitude from port wing shaker (PW\_LC) to PW\_ULE3 during the panel removal tests (amplitude level 5).}
	    \label{fig:panel_PW}
	\end{figure}
	
	As previously stated, the principal purpose of the Hawk dataset is to provide a meaningful benchmark for \ac{shm} in aerospace structures. It is intended that the presented dataset is able to function as a useful tool to bridge the gap between small scale laboratory experiments and complete operational systems. The level of system complexity seen in this dataset greatly complicates the application of many \ac{shm} approaches. Opportunities exist to extend existing and develop new techniques spanning the all levels of Rytter's hierarchy barring remaining life prognosis \cite{phd:Ryt:1993}.
	
	Various levels and types of (pseudo-)damage have been simulated to provide a number of scenarios under which methods may be tested. The addition of masses has been conducted in the same manner as shown in \cite{art:Hay:2024}, this point is important as it allows the evaluation of methods from a \ac{pbshm} viewpoint where transfer learning may be applied between the two alternative sensing and excitation setups. The effects of these added mass simulated damages are illustrated in Figures \ref{fig:dmg_PW} and \ref{fig:dmg_SS} for the cases where damage has been simulated on the port wing and starboard stabiliser respectively. Both figures show pronounced effects at higher levels of added mass (mass references in figure legends refer to Table \ref{tab:masses}). It is expected that detection of these damage scenarios should be achievable on the basis of the response in the higher frequency ranges.
	
	
	Of particular interest is the effectiveness of methods using only the strain data or methods which seek to perform classification or estimation of damage extent. It would be especially desirable if these quantities could be transferred between this test dataset and that previously published on only a single wing \cite{art:Hay:2024}. 
	
	A further challenge is the case of panel removal. Such a case is shown in Figure \ref{fig:panel_PW}, where \ac{frf}s are shown for panels removed from the port wing of the aircraft. The effect of the panel removals is, at least optically, significantly harder to detect than that of the added masses and should be considered as a more meaningful challenge with respect to the testing of any \ac{shm} methodology.
	
	If \ac{shm} technologies are to see widespread adoption, they must consider whether it is practical to collect required data for different algorithms during the operational life of the structure. There exists a distinction between what data would be available in a laboratory or development setting and what could be practically collected in service. Installing a comprehensive network of accelerometers on an aircraft for use in flight is a concept that presents many practical difficulties. It is for this reason that \ac{fbg} data was included in this dataset, in addition to the accelerometer data.  Alternative sensing techniques, such as \ac{fbg} sensors, may provide more viable routes to in-service monitoring of structural vibrations. However, each sensing modality comes with its own restrictions and limitations. For example, in the case of the \ac{fbg} data, a much lower signal-to-noise ratio is present, which may be observed when comparing Figure \ref{fig:all_strain} to Figure \ref{fig:all_accs}. Performing \ac{shm} with this sparser, noisier data presents a greater challenge than with the accelerometer data.
	
	
	Finally, an open question remains concerning the change in available data between the ``training'' phase of an \ac{shm} algorithm and that which is used in ``testing'' (i.e. practical usage). This can be summarised as `Should ``higher quality'' data be used during algorithmic development with a view to adapt to the data available in use or should only operationally available data be used?'. The picture is complicated even further if a population-based approach \cite{gardner2021foundations} is taken. It is hoped that the data set presented here can be used to further explore this question in the future. 

	Dataset challenges in \ac{shm} are summarised below. 
	
	
	\begin{tcolorbox}
	    \begin{enumerate}[label={Challenge \arabic*.}, leftmargin=3cm, listparindent=-\leftmargin, resume*]
	        \item{Detecting and classifying pseudo-damage using the entirety of the available data.}
	        \item {Detecting and classifying pseudo-damage using subsets of the available data.}
	        \item {Detecting and classifying panel removal with any amount of the available data.}
	        \item {Detecting and classifying pseudo-damage with exclusively the FBG data.}
	        \item {Establishing the extent to which higher quality data is useful in the training phase, when only lower quality data is available during operation.}
	    \end{enumerate}
	\end{tcolorbox}

\section{Conclusions}

	This paper presents a new comprehensive dataset for \ac{shm}, acquired from a BAE Systems Hawk T1A. The intention of the dataset is to provide a benchmark for the development of vibration-based \ac{shm} and \ac{stid} methods on a full-scale aerospace structure, with the benefit of laboratory testing conditions.
	
	The key details of the experimental work carried out to acquire the data have been described. Representative sections of data have been presented to highlight damage sensitivity, non-linearity and strain responses of the structure. These examples were presented to indicate the potential applications and challenges of the dataset in future research work.
	
	The key challenges that are foreseen based on this dataset relate primarily to \ac{stid} and \ac{shm}. Here, the large scale of the dataset, the significant complexity in the dynamics (including non-linearity) and the use of strain measurements were highlighted as future challenges for the research community to address.

\section*{Data availability}

	All data collected is made freely available at: \url{orda.shef.ac.uk/articles/dataset/BAE_T1A_Hawk_Full_Structure_Modal_Test/24948549}
	
	In addition, the authors make a Python interface for interacting with the data available at: \url{https://github.com/MDCHAMP/hawk-data}

\section*{Author contribution CRediT statement}

	JW: Methodology (Equal), Investigation (Lead), Writing - Original Draft (Equal), Writing - Review \& Editing (Equal)
	
	MDC: Software (Lead), Data Curation (Lead), Writing - Original Draft (Equal), Writing - Review \& Editing (Equal), Visualisation (Lead)
	
	MT: Methodology (Equal), Investigation (Lead), Writing - Original Draft (Equal), Writing - Review \& Editing (Equal), Visualisation (Lead)
	
	RM: Methodology (Equal), Investigation (Supporting)
	
	DW: Conceptualisation (Lead), Resources (Lead), Writing - Review \& Editing (Equal), Supervision (Lead), Funding acquisition (Lead)
	
	TJR: Conceptualisation (Lead), Resources (Lead), Writing - Original Draft (Supporting), Writing - Review \& Editing (Equal), Supervision (Lead), Funding acquisition (Lead)

\section*{Acknowledgements}

	This work was funded by the Alan Turing Institute Research and Innovation Cluster for Digital Twins and EPSRC grant EP/R006768/1. The authors gratefully acknowledge this support. MDC would additionally like to acknowledge the support of EPSRC grant EP/W005816/1.
	
	Finally, the authors extend their thanks to Adam Brassington, Daniel Clarkson, Iori Fukuda and Collins Ogbodo for their assistance in collecting the data.

\bibliographystyle{elsarticle-num}
\bibliography{Hawk_references}

\end{document}